\documentclass[10pt]{iopart}
\usepackage{iopams, graphicx,url, bm, color, tikz, pgfplots, hyperref, times, xspace}
\usepackage{siunitx}[=v2]
\expandafter\let\csname equation*\endcsname\relax
\expandafter\let\csname endequation*\endcsname\relax
\usepackage{amsmath, amsfonts, amssymb}
\usepackage{chemmacros}
\usepackage[switch]{lineno}
\usepackage[backend=biber, sorting=none, bibencoding=utf8, url=false, doi=true, eprint=false, isbn=false, firstinits, maxcitenames=2, maxbibnames=3]{biblatex}

\hypersetup{colorlinks=true, linkcolor=blue, citecolor=blue, urlcolor=blue}
\providecommand\fpos{\ensuremath{\text{1PN2}}}
\providecommand\e{\ch{e}}
\providecommand\otwo{\ch{O2}}

\DeclareSourcemap{
  \maps[datatype=bibtex]{
    \map{
      \step[fieldset=abstract, null]
    }
  }
}

\DeclareSIUnit{\townsend}{Td}

\begin{document}

\title{Genesis of column sprites: Formation mechanisms and optical structures}

\author{R Marskar}
\address{SINTEF Energy Research, Sem S\ae lands vei 11, 7034 Trondheim, Norway.}
\ead{robert.marskar@sintef.no}

\begin{abstract}
  Sprite discharges are electrical discharges that initiate from the lower ionosphere during intense lightning storms, manifesting themselves optically as flashes of light that last a few milliseconds. 
  This study unravels sprite initiation mechanisms and evolution into distinctive morphologies like glows and beads, using direct 3D numerical simulations that capture the intricate electrical discharge processes.
  We clarify various morphological aspects of sprites such as the halo dynamics, column glows, branching, streamer reconnection, and bead formation. 
  The results advance our understanding of sprites and their connection to thunderstorm dynamics, and puts quantitative analysis of their effect on Earth's climate within reach.  
\end{abstract}

\submitto{\PSST}

\maketitle

\ioptwocol

\section{Introduction}
The phenomenon known as sprite discharges in the upper atmosphere has captivated scientists and skygazers for decades.
Often referred to as red sprites due to their distinctive reddish glow, they manifest as millisecond-short flashes of light high above thunderstorms (\SIrange{40}{80}{\kilo\meter}).
The initial inklings of sprite discharges trace back hundreds of years to eyewitnesses reports, while more recent observations have been reported by amateur photographers, pilots, and astronauts.
These anecdotal accounts were validated when the first documented evidence was conclusively captured on camera in 1989 \cite{Franz1990}.
Since then, high-speed photography \cite{Cummer2006,McHarg2008} have unveiled a rich tapestry of sprite activity, establishing sprites as a field of study that offers unique insights into the electrical interactions within Earth's atmosphere.
Due to advances in consumer-grade hardware, sprites are now routinely captured also by amateur photographers, as exemplified by the NASA 'Spritacular' citizen science project \cite{Kosar2022}.

Research on sprites transcends scientific curiosity as they emit bursts of radio waves and disrupt the ionosphere, influence the propagation of radio signals, and introduce uncertainties in communication and navigation systems.
Furthermore, sprites can generate reactive species in the Earth atmosphere in large quantities and thus affect Earth's long term atmospheric composition.
Unraveling the nature of sprite streamers thus holds significant implications for understanding Earth's atmospheric electricity, ionospheric dynamics, and climate. 

It is now accepted that sprites originate during intense lightning discharges from thunderstorms, which generate quasi-static electric fields that extend into the upper atmosphere.
As the breakdown strength $E_{\text{br}}$ of air relative to the number density of neutral molecules $N$ is a constant (i.e., $E_{\text{br}}/N \approx \SI{120}{Td}$ in air), and $N$ decreases exponentially with altitude, powerful lightning strikes can initiate electrical breakdown in the upper atmosphere \cite{Wilson1924}.
The quasi-static electric fields that originate in charged thunderclouds do not penetrate the ionosphere as it is sufficiently conductive and, analogous to metals, has no internal electric field on the time scale of sprites.
But the fields initiate a cascade of processes at the lower edge of the ionosphere which ultimately lead to the formation a filamentary type of plasma known as streamers \cite{Nijdam2020}.
These structures propagate through self-enhanced electric fields at their tips, and can propagate downwards in the atmosphere for tens of kilometers.
The optical emission from sprites result from electron impact excitation and spontaneous emission from the first and second positive system of molecular nitrogen (1PN2 and 2PN2), resulting in the display of red and purple hues that often define sprites.

\begin{figure}[h!t!b!]
  \centering
  \includegraphics[width=0.4\textwidth]{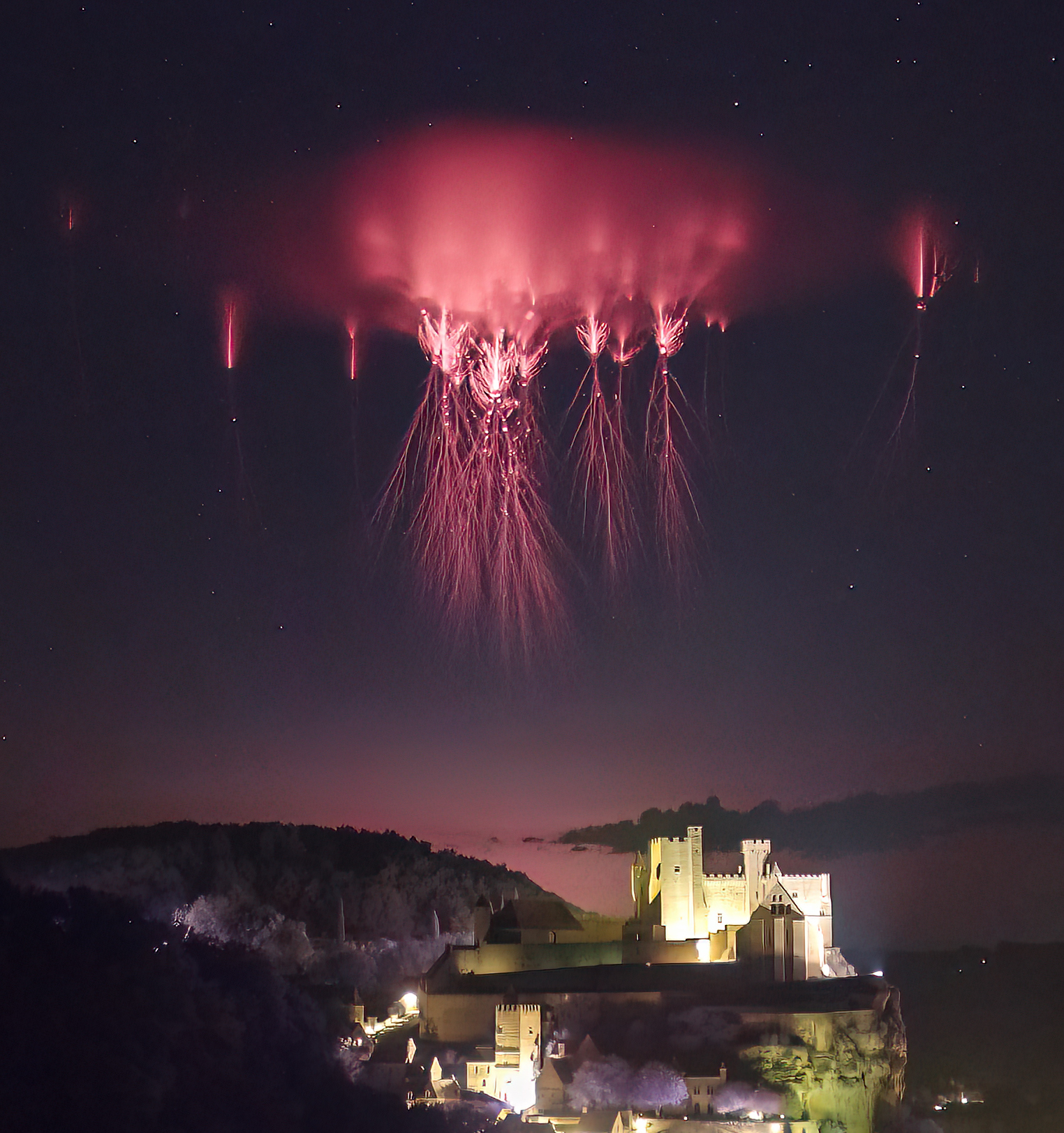}  
  \caption{
    True color photo of a large carrot sprite over France.
    Courtesy of Nicolas Escurat via the Spritacular citizen project \cite{Kosar2022}.
  }
  \label{fig:RedSprite}
\end{figure}

Most sprite discharges occur during positive cloud-to-ground (+CG) lightning and display a fascinatingly rich morphology, see \fref{fig:RedSprite}.
High-speed video studies \cite{Stanley1999} show that sprites are often preceded by a sprite halo \cite{Wescott2001, BarringtonLeigh2001,Miyasato2002}, which is a large diffuse glow with lateral extents exceeding tens of kilometers.
Although halos from +CG and -CG lightning occur with almost the same frequency for a given lightning current \cite{Newsome2010}, virtually all sprites arise from +CG lightning, while negative sprites are rarely seen \cite{Williams2007}.
Sprites are classified as column or carrot sprites, with the primary difference being that column sprites consist of only downwards propagating streamers and carrot sprites also consist of upwards propagating streamers.
Telescopic imaging \cite{Gerken2000, Marshall2005, Marshall2006} shows that sprite discharges have a fine structure consisting of long-lasting column-shaped glows near the top of the sprite, and occasionally also spots (i.e., beads) along the streamer channels.
This optical fine structure dominates long-exposure images of sprites, while the streamer channels leave relatively faint traces, as shown in \fref{fig:RedSprite}.

While sprite discharges have been a subject of scientific scrutiny since their indisputable discovery 35 years ago, many unanswered questions remain.
One key mystery pertains to the precise mechanism triggering the formation and morphological structure of sprites \cite{Luque2009}.
It is widely accepted that they are associated with lightning strikes, but the specific conditions and processes leading to their creation remain incompletely understood, as are the intricate details of sprites' complex internal structure and behavior.
This includes factors that determine their distinctive shapes, sizes, and morphology, such as the breakup of the halo, and the appearance of long-lasting columniform glows and luminous beads.
Many of these questions have remained theoretically inaccessible, primarily owing to a lack of appropriate theoretical tools that capture the full three-dimensional evolution of the sprite across the many time and length scales that are involved in their formation.

In this paper we theoretically study, for the first time, the complete three-dimensional formation of a moderately-sized atmospheric column sprite with realistic theoretical models and atmospheric conditions.
We track the evolution of the sprite from its inaugural birth in the sprite halo and as it descends in the atmosphere.
Our results reproduce many optical signatures associated with sprites, such as 1) the appearance and breakup of an initial halo, 2) downwards shooting positive streamers, 3) streamer branching, 4) sprite glows, 5) streamer reconnections, and 6) sprite beads.
We also find that these processes have rather clear physical interpretations and emergence mechanisms.

This paper is organized as follows:
In \sref{sec:model} we discuss the physical and computer models that are used.
In \sref{sec:results} we provide observations of the various stages of the sprite, and provide step-by-step interpretations of their physical mechanisms.
We compare our observations with past modeling studies as well as experimental observations wherever possible.
Finally, in \sref{sec:conclusion} we provide a brief summary of the paper.

\section{Theoretical model}
\label{sec:model}

\subsection{Plasma chemistry and evolution}
\begin{figure}[htb]
  \centering
  \includegraphics{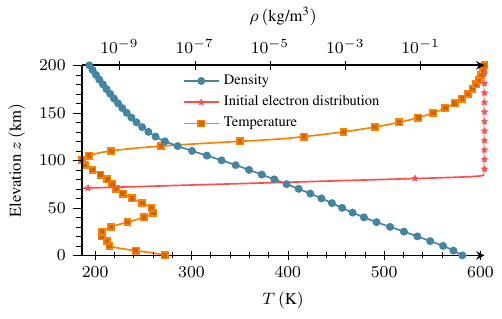}
  \caption{Electron density profile (arbitrary units), and atmospheric gas density and temperature as a function of elevation $z$.
  The atmospheric density was computed with an ENMSIS atmospheric model.}
  \label{fig:Atmosphere}
\end{figure}

We model the plasma as densities using 3D advection-diffusion-reaction equations for species number densities $n_i$,

\begin{equation}
  \partial_t n_i + \nabla\cdot\left(\bm{v}_i n_i - D_i\nabla n_i\right) = S_i,
\end{equation}
where $i$ is some species (e.g., electrons or a type of ion), $\bm{v}_i = \pm \mu_i\bm{E}$ is the advective velocity for species with positive (+) and negative (-) charge numbers, $\mu_i$ are mobilities, $D_i$ are diffusion coefficients, and $S_i$ are source terms.
The electric field $\bm{E}$ is obtained from the Poisson equation

\begin{equation}
  \nabla\cdot{\bm{E}} = \frac{\rho}{\epsilon_0},
\end{equation}
where $\rho = \sum_i q_{\text{e}}Z_in_i$ where $q_{\text{e}}$ is the elementary charge and $Z_i$ are charge numbers.
For greatest simplicity, the Earth atmosphere is modeled as a gaseous background of \SI{20}{\percent} \ch{O2} and \SI{80}{\percent} \ch{N2}, where the temperature $T$ and density $\rho$ are given as functions of altitude $z$ above sea-level as shown in \fref{fig:Atmosphere}.
We solve the above equations self-consistently for electrons $\ch{e}$, ions $\ch{N2^+}$, $\ch{O2^+}$, and $\ch{O^-}$, while the relative change in the densities of $\ch{O2}$ and $\ch{N2}$ are ignored as the ionization degree is typically below \SI{E-4}{}.
Chemical reactions are given in \tref{tab:reactions}, and transport data for the electrons are computed using BOLSIG+ \cite{Hagelaar2005} and the SIGLO database \cite{SigloDB}.
Since the transport data only shows minor variations with temperature for \SIrange{200}{300}{\kelvin}, it is computed for a constant temperature $T=\SI{200}{\kelvin}$.
Ions are considered as stationary.

\begin{table*}[htb]
  \caption{
    \label{tab:reactions}
    Simplified sprite chemistry used in the model. \ch{N2} and \ch{O2} indicate ground states, while optical emission follows from the first positive system $\ch{N2}\left(\ch{B^3}\Pi_g\right) \rightarrow \ch{N2}$.
  }
  \begin{indented}
  \item[]
    \begin{tabular}{lllll}
      \hline
      \# & Type & Reaction & Symbol & Reference \\
      \hline      
      $\text{R}_1$      & Ionization& $\ch{e} + \ch{N2} \rightarrow \ch{e} + \ch{e} + \ch{N2^+}$  & $k_1(E,z)$ & \cite{Hagelaar2005, SigloDB}\\
      $\text{R}_2$      &      Ionization& $\ch{e} + \ch{O2} \rightarrow \ch{e} + \ch{e} + \ch{O2^+}$  & $k_2(E,z)$ & \cite{Hagelaar2005, SigloDB}\\
      $\text{R}_3$      &      Dissociative attachment& $\ch{e} + \ch{O2} \rightarrow \ch{O} + \ch{O^-}$  & $k_3(E,z)$ & \cite{Hagelaar2005, SigloDB}\\
      $\text{R}_4$      &      Optical emission& $\ch{e} + \ch{N2} \rightarrow \ch{e} + \ch{N2}\left(\ch{B^3}\Pi_g\right)$  & $k_4(E,z)$ & \cite{Hagelaar2005, SigloDB, Liu2004}\\
      $\text{R}_\gamma$ &      Photon production& $\ch{e} + \ch{N2} \rightarrow \ch{e} + \ch{N2} + \gamma$  & $k_\gamma(E,z)$ & \cite{Hagelaar2005, SigloDB, Pancheshnyi2005}\\
      \hline
    \end{tabular}
  \end{indented}  
\end{table*}

\subsection{Photoionization}

Photoionization is treated using Monte Carlo sampling as in \cite{Bagheri_2019,Marskar2020}, corrected for altitude.
For \ch{N2}-\ch{O2} mixtures, photoionization of molecular oxygen can occur by electron impact excitation of nitrogen to one of the involved singlet states (denoted by \ch{N2^{exc}} below) \cite{Liu2004,Stephens2018},

\begin{equation}
  \ch{e} + \ch{N_2} \rightarrow \ch{e} + \ch{N_2^{exc}}.
\end{equation}
If \ch{N_2^{exc}} is not predissociated or quenched, the state decays through spontaneous emission $\ch{N2^{exc}} \rightarrow \ch{N2} + \gamma$ which can lead to photoionization of molecular oxygen:

\begin{equation}
  \gamma + \ch{O2} \rightarrow \ch{e} + \ch{O2^+}.
\end{equation}
The physical number of photons that are generated per unit time and volume is

\begin{equation}
  S_\gamma = k_\gamma n_{\ch{e}} n_{\ch{N2}}
\end{equation}
where

\begin{equation}
  k_\gamma = \frac{p_q}{p + p_q}\nu_Z(E)k_1.
\end{equation}
Here, $p = p(z)$ is the gas pressure and $p_q=\SI{40}{\milli\bar}$ is a quenching pressure which describes collisional de-excitation of the various \ch{N2} singlet states \cite{Stephens2018}.
This factor is included, although we point out that the quenching provides a negligible correction at sprite altitudes, but reduces photoionization by approximately a factor of \num{0.04} at atmospheric pressure.
The function $\nu_Z(E)$ is a correction factor that accounts for photoionization probabilities and excitation efficiencies, see \textcite{Pancheshnyi2015}.
The physical number of photons that is generated in each time step $\Delta t$ in a grid cell with volume $\Delta V$ is then sampled as $P\left(S_\gamma \Delta t \Delta V\right)$, where $P\left(\mu\right)$ is a Poisson distribution with mean value $\mu$.
When computational photons are generated we sample the mean absorption length $\kappa_f$ individually for computational photons with frequency $f$ using

\begin{equation}
  \kappa_f = p_{\ch{O2}}\chi_{\text{min}} \left(\frac{\chi_{\text{max}}}{\chi_{\text{min}}}\right)^{\frac{f-f1}{f2-f1}},
\end{equation}
where $p_{\otwo}$ is the partial pressure of oxygen, $\chi_{\text{min}} = \SI[per-mode=reciprocal]{2.625E-2}{\per\pascal\per\meter}$ and $\chi_{\text{max}} = \SI[per-mode=reciprocal]{1.5}{\per\pascal\per\meter}$.
The photon frequency $f$ is uniformly sampled on the interval $f\in[f_1,f_2]$ where $f_1=\SI{2.91}{\peta\hertz}$ and $f_2=\SI{3.06}{\peta\hertz}$ \cite{Liu2004}.

\begin{figure}[htb]
  \includegraphics{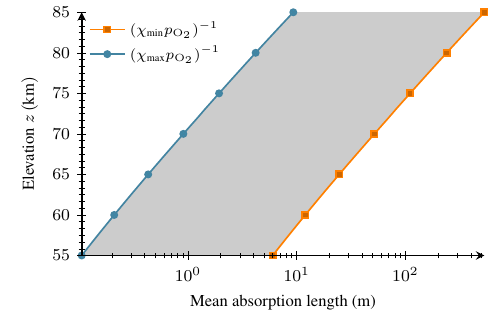}
  \centering
  \caption{Minimum and maximum mean absorption lengths for ionizing photons at various altitudes.
    The mean absorption length of all ionizing photons in our model falls inside the shaded region.}
  \label{fig:AbsorptionLength}
\end{figure}

\Fref{fig:AbsorptionLength} shows the calculated maximum and minimum mean absorption lengths in our photoionization model as a function of altitude.
In the computer implementation the absorption position of a computational photon is doubly sampled: We first sample $f$ on the specified interval and compute $\kappa_f$ for each photon, and then sample the propagation distance of photons by sampling an exponential distribution with parameter $\kappa_f$.
We limit the photon generation to a maximum of 32 computational photons per cell per time step, which introduces some additional stochastic fluctuations in the ionization level ahead of the streamers.

\subsection{Optical signature}
\label{sec:OpticalSignature}
For comparison with optical observations we calculate the number of excitations into the first positive system $\ch{N2}\left(\ch{B^3}\Pi_g\right)$ of molecular nitrogen.
We represent the accumulated optical emission density until time $t$ as

\begin{equation}
  \label{eq:1PN2Phi}  
  n_{\text{1PN2}} = \frac{A_k}{A_k + k_q N} n_{\ch{N2}\left(\ch{B^3}\Pi_g\right)},
\end{equation}
where the spontaneous emission rate is $A_k=\SI[per-mode=reciprocal]{1.7E5}{\per\second}$ and the collisional quenching constant is $k_q=\SI{8E-18}{\cubic\meter\per\second}$ \cite{Liu2004}, which compensates for reduced optical emissions at lower altitudes.
The relative radiative emission rate $\frac{A_k}{A_k + k_q N}$ from $\ch{N2}\left(\ch{B^3}\Pi_g\right)$ varies from approximately \num{0.99} at $z=\SI{80}{\kilo\meter}$ to \num{0.8} at $z=\SI{60}{\kilo\meter}$.

The effective lifetime of the $\ch{N2}\left(\ch{B^3}\Pi_g\right)$ state is $A_k^{-1} \approx \SI{6}{\micro\second}$, while typical time scales for the morphological evolution of sprites are \SIrange{E-5}{E-3}{\second}.
Excitation into and emission from the $\ch{N2}\left(\ch{B^3}\Pi_g\right)$ state is thus virtually instantaneous on the sprite time scale.
As with the accumulated emission, the instantaneous emission is represented as

\begin{equation}
  \label{eq:1PN2Src}
  S_{\fpos} = \frac{A_k}{A_k + k_q N} S_{\ch{N2}\left(\ch{B^3}\Pi_g\right)}.
\end{equation}

\subsection{Initial data}
We consider an initial electron density

\begin{equation}
  \label{eq:init_ne}
  n_{\text{e}}(z) = \frac{n_0}{2}\left[1+\tanh\left(\frac{z-z_0}{L}\right)\right],
\end{equation}
with $n_0 = \SI[per-mode=reciprocal]{5E9}{\per\cubic\meter}$, $z_0=\SI{83}{\kilo\meter}$ and $L = \SI{1}{\kilo\meter}$, i.e. with negligible charge except in the ionosphere where $n_{\text{e}}\approx \SI[per-mode=reciprocal]{5E9}{\per\cubic\meter}$.
The initial ion densities are $n_{\ch{N2^+}} = 0.8n_{\ch{e}}$ and $n_{\ch{O2^+}} = 0.2n_{\ch{e}}$, respectively.

The thundercloud field is modeled as an initially electrically neutral cloud which transfers charge to the ground such that the space charge density in the thundercloud is

\begin{equation}
  \rho_{\text{tc}}\left(t,\bm{x}\right) = -\rho_0\exp\left[-\frac{1}{2}\left(\frac{\left|\bm{x}-\bm{x}_0\right|}{R}\right)^4\right]\left(1 - \text{e}^{-t/\tau}\right),
\end{equation}
with $R=\SI{10}{\kilo\meter}$, $\bm{x}_0$ set to \SI{10}{\kilo\meter} above ground, $\tau = \SI{0.1}{\milli\second}$, and $\rho_0=\SI{6.4E-3}{\coulomb\per\cubic\kilo\meter}$.
The total charge in the cloud is approximately \SI{-41}{\coulomb}.

No other sources of preionization \cite{Kohn2019} or preconstructed field distributions \cite{Malagon2020} were used in order to trigger any specific features of the sprite.
Our model therefore permits the sprite to evolve naturally from only the ionospheric electrons and the electric field imposed by the charge distribution in the thundercloud.

\subsection{Numerical discretization}
All calculations in this paper were done using the chombo-discharge code \cite{Marskar2023}, which we have used for numerous discharge simulations in the past \cite{Marskar2019a,Marskar2019b,8785920,Meyer2020,Marskar2020,Meyer2022,Marskar2023}.
We solve the equations of motion over a $\SI{400}{\kilo\meter}\times\SI{400}{\kilo\meter}\times\SI{200}{\kilo\meter}$ region, using an adaptive Cartesian mesh.
The base grid consists of $256\times256\times128$ cells, with another 11 levels of mesh refinement that are dynamically adapted throughout the simulation, i.e., the effective grid size is $\num{524288}\times\num{524288}\times\num{262144}$ cells and the finest representable grid resolution is approximately \SI{0.76}{\meter}.
Grid cells are refined if

\begin{equation}
  \overline{\alpha}\Delta x \geq 1,
\end{equation}
and coarsened if

\begin{equation}
  \overline{\alpha}\Delta x \leq 0.2,
\end{equation}
where $\overline{\alpha}$ is the effective Townsend ionization coefficient and $\Delta x$ is the local grid resolution.

The temporal integration uses a Godunov splitting between plasma transport and reactions, where the transport equations are discretized in time using a Corner Transport Upwind (CTU) scheme with a monotonized central limiter \cite{Colella1990} for the normal slopes.
The Poisson equation is solved using a geometric multigrid method with V-cycling.
We use a semi-implicit coupling between the Poisson equation and the transport equations (e.g., see \cite{Ventzek1994}) in order to eliminate the dielectric relaxation time step contraint.
Complete details regarding the discretization method are given in the chombo-discharge documentation \cite{Marskar2023}.
The simulations in this paper used up to \num{25600} CPU cores with mesh sizes up to \num{3E9} grid cells, and completed in about a day.
Radiative transfer accounted for about \SI{40}{\percent} of a time step cost, which is due to sampling of approximately \num{20E9} computational photons per time step.
Unfortunately, we were unable to load balance our way out this bottleneck as the remapping of photons to CPU ranks took place as an all-to-all process using a substantial amount of computational particles, which led to some load imbalance in the communication pattern for the discrete photons.

\begin{figure}[h!t!b!]
  \centering
  \includegraphics{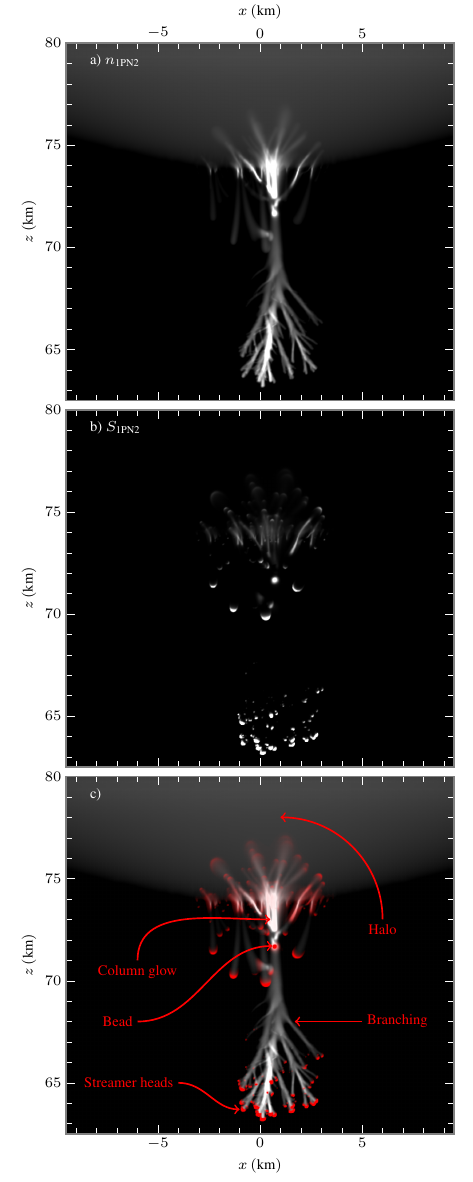}
  \caption{
    Images of the optical emissions from a sprite discharge simulation.
    a) Accumulated emissions.
    b) Instantaneous emission.
    c) Overlaid emissions.
  }
  \label{fig:SpriteEmission}
\end{figure}

\section{Results}
\label{sec:results}
Our results track a sprite from its birth in the halo after the lightning strike until $t\approx\SI{11}{\milli\second}$.
The sprite initiated at approximately $z=\SI{80}{\kilo\meter}$ and extended downwards to approximately \SI{63}{\kilo\meter}, at which point we terminated the computer simulation due to a decreasing reliability as the numerical resolution relative to the physical quantities of interest decreased.
When we analyze the computer simulations we consider various quantifications such as three-dimensional isosurface reconstruction, planar slices, and images.
The latter are defined by solving a simplified radiative transfer problem between the sprite emissions ($n_{\fpos}$ and $S_{\fpos}$) and a reference plane that looks into the sprite from either the $x$ or $y$ direction.
Alpha channels are then added to the instantaneous emission images, which are then overlaid the images of the long-exposure emissions.
\Fref{fig:SpriteEmission} shows an image of the sprite emissions at $t=\SI{10.72}{\milli\second}$ along with several indicated optical signatures, and is the simulation case that we systematically analyze.

\subsection{Halo initiation}
\label{sec:HaloInitiation}
\begin{figure}[htb]
  \centering
  \includegraphics{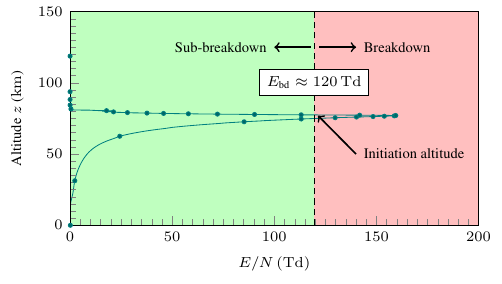}
  \caption{Reduced electric field above the thundercloud after the initial lightning strike ($t=\SI{1.25}{\milli\second}$).}
  \label{fig:FieldDropoff}
\end{figure}

First, we discuss initiation of the halo.
Charge transfer from cloud to ground results in a dipole field from the remaining positive thundercloud charge and ground, which drops off with altitude as $z^{-1}$.
As the atmospheric density decays exponentially with altitude it results in a field $E/N > E_{\text{br}}$ at altitudes \SIrange{75}{83}{\kilo\meter} after $t\approx \SI{0.15}{\milli\second}$, which is where the halo initiates.
Initiation does not extend into the ionosphere since the dielectric relaxation time is $\mathcal{O}\left(\SI{E-10}{\second}\right)$ in this region, and thus the electric field is negligible above altitudes of \SI{83}{\kilo\meter}.
\Fref{fig:FieldDropoff} shows how the reduced electric field has increased above the thunderstorm after the lightning strike, showing a thin layer of above-breakdown fields at altitudes of $z\approx\SI{80}{\kilo\meter}$.

\begin{figure}[htb]
  \includegraphics{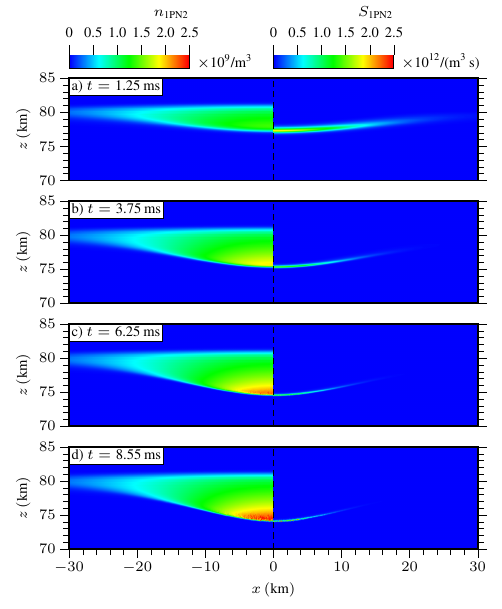}  
  \caption{
    Cross-section of the simulation data showing the accumulated ($n_{\fpos}$) and instantaneous ($S_{\fpos}$) optical emissions.
    The data shows slices through mesh-based densities at the center of the sprite halo (i.e., they are not projections of emissions onto a camera aperture.)
  }
  \label{fig:HaloInitiation}
\end{figure}

The time evolution of the halo emissions is given in \fref{fig:HaloInitiation}, showing data-slices of $n_{\fpos}$ and $S_{\fpos}$ through the center of the halo.
Optically, this will be observed as a large, luminous region that propagates downwards from the lower ionosphere, which is consistent with high-speed imaging observations \cite{Cummer2006} as well as several other theoretical studies \cite{Luque2009}.

We have not studied the initiation altitude of the sprite halo with other parameters, but observe that this is determined by 1) the charge moment (and temporal characteristic) of the lightning strike, and 2) the altitude of the ionospheric boundary.
The ionosphere has no internal electric field on the timescale of sprites, so its lower boundary determines the upper cut-off of $E/N$ as shown in \fref{fig:FieldDropoff}.
As $N$ decays exponentially with altitude, the precise altitude of the ionospheric boundary is of some importance since deviations by just a few kilometers lead to large variations in the charge moment required for initiating the halo.

\subsection{Halo breakup and streamer formation}
\label{sec:HaloBreakup}
\begin{figure}[htb]
  \centering
  \includegraphics{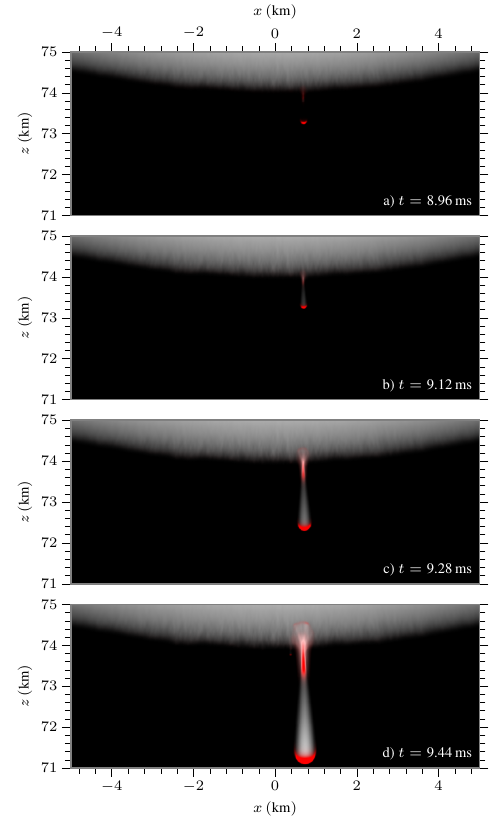}
  \caption{
    Accumulated and instantaneous optical emission images from $n_{\fpos}$ during halo breakup.
    The emerging streamer emits light both from its tail and its downwards propagating head.
  }
  \label{fig:HaloBreakup}
\end{figure}

Next, we present computational observations preceding halo breakup into streamer filaments.
Experimental high-speed imaging \cite{Cummer2006} shows that downward propagating streamers initiate at bottom of the halo, either spontaneously or from brightening mesoscopic inhomogeneities.
\Fref{fig:HaloBreakup} shows the formation of such a streamer in the computer simulations, which here initiates at the bottom of the halo at $z\approx\SI{74}{\kilo\meter}$ after $t\approx \SI{8}{\milli\second}$.
Initially, only a single columniform streamer emerges, whose optical signature is characterized by emission from the streamer head and a column glow in its tail.
Other streamers emerge later from the halo, but are not yet discernible in \fref{fig:HaloBreakup}.
We also observe that streamer emergence from the halo is not instantaneous and there is a significant temporal delay between the stagnation of the halo and the emergence of the sprite.
\Fref{fig:HaloInitiation} shows that the halo front reaches an altitude of $z\approx \SI{75}{\kilo\meter}$ after $t=\SI{6.25}{\milli\second}$, but the streamer emerges about two milliseconds later.

Various mechanisms have been suggested as explanations for the halo breakup, such as ''collapse'' of the halo into a sprite streamer \cite{Luque2009}, or initiation through mesospheric disturbances such as gravity waves \cite{Liu2015}.
Here, a single streamer emerges spontaneously from the halo, which we attribute to stochastic fluctuations in the electron density below the halo edge.
These fluctuations appear because as the halo descends downwards it also encounters an increasing air density, which approximately doubles every \SI{5}{\kilo\meter}.
The production of ionizing photons at the halo edge then decreases, as does the mean free path length of the photons.
Photoionization in front of positive streamers is known to stabilize filaments \cite{Bagheri_2019, Marskar2020} in the sense that as photoionization is increased, electron density fluctuations in front of the streamers decrease.
These fluctuations would otherwise grow exponentially towards the streamer head (or halo edge, in this case).
Correspondingly, as photoionization is reduced these density fluctuations become increasingly relevant.

\begin{figure}[htb]
  \centering
  \includegraphics{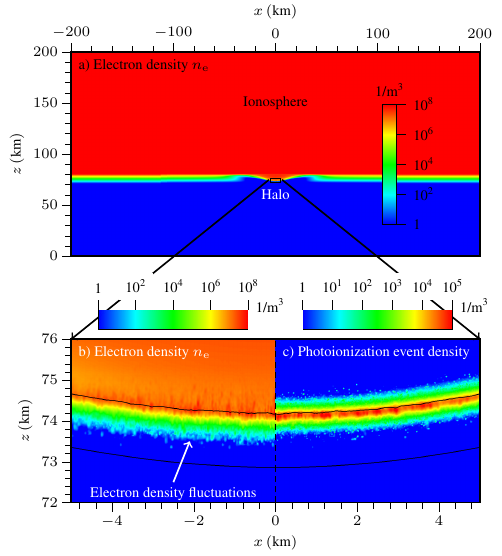}
  \caption{Electron density and photoionization distribution at the lower halo edge preceding halo breakup at $t=\SI{8.57}{\milli\second}$.
    The photoionization data shows the number of photoionization events per unit volume over an averaging time of approximately \SI{1}{\micro\second}.
    The curved black lines in the lower two panels show contours $\alpha=\eta$.
  }
  \label{fig:HaloFluctuations}
\end{figure}

\Fref{fig:HaloFluctuations} shows the electron density and photoionization density near the halo edge, plotted together with a contour of $\alpha=\eta$ which shows the ionization zone.
We observe, firstly, that the ionization zone is much longer ($\approx \SI{1.3}{\kilo\meter}$) than the photon absorption length (\SIrange{1}{100}{\meter}) and few photons (if any) reach the edge of the ionization zone.
Substantial fluctuations in the electron density in the ionization zone therefore occur as most photons are absorbed very close to the halo front.
A comparatively low number of photons travel further into the ionization zone.

The fluctuations imposed by photoionization are superimposed on top of the ionospheric electrons.
However, \fref{fig:HaloFluctuations}b shows that the electron density ahead of the halo at $z=\SI{73}{\kilo\meter}$ is $n_\e \approx \SI[per-mode=reciprocal]{0}{\per\cubic\meter}$, while equation~\eref{eq:init_ne} indicates a value of $n_\e \approx \SI[per-mode=reciprocal]{10}{\per\cubic\meter}$.
Similarly, following the electron density in \fref{fig:HaloFluctuations}a) radially outwards from the center of the halo, e.g., on $z\approx \SI{75}{\kilo\meter}$, we observe that the electron density has decreased to negligible levels on the sides of the halo.
This decrease in electron density, both ahead of the halo as well as on its sides, occurs due dissociative attachment where free electrons are rapidly converted into negative ions during the halo descent.
A more thorough discussion of dissociative attachment follows later in the paper, but we point out that at the halo initiation altitude this conversion occurs on the order of a few milliseconds (see \fref{fig:RateCoefficients}), which implies that the number of free electrons ahead of the halo rapidly decreases after the lightning strike. 
The main implication is that the preionization ahead of the ionization zone of the halo is exponentially reduced in time, and correspondingly that the ambient ionospheric electrons ahead of the halo do not have a major impact on the magnitude of the photoionization-induced fluctuations.

\begin{figure}[htb]
  \centering
  \includegraphics{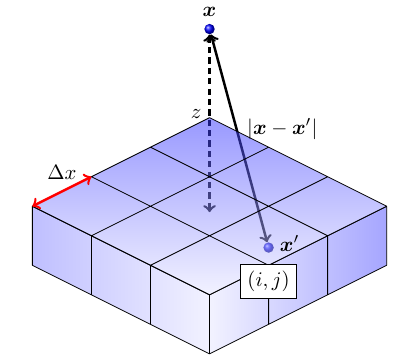}
  \caption{Toy model for photon emission and absorption from a planar front of grid cells.
    Each cubic cell represents a numerical grid cell and isotropically emits $N_\gamma$ physical photons.
  }
  \label{fig:PlanarFront}
\end{figure}

The role of photoionization on the branching of positive and negative streamers have earlier been given by \textcite{Liu2004}.
Here, we focus on the onset of discrete photon effects.
One may derive an estimate for the distance at which the discrete nature of photons becomes non-negligible during a computational time step, using a simplistic model for the emission and absorption of ionizing photons.
We consider a horizontal slice through the descending halo, which can be simplistically represented as a planar front consisting of neighboring grid cells with spatial extents $\Delta x$, as indicated in \fref{fig:PlanarFront}.
In this figure the location of each grid cell is $\bm{x}^\prime = i\Delta x\bm{\hat{x}} + i\Delta x\bm{\hat{y}}$, and we assume that each grid cell isotropically emits $N_{\gamma}$ photons.
Note that the toy model coordinates are different from the sprite simulation coordinates. 
The average number of photons that are absorbed per unit volume at position $\bm{x}$ from the emission site $\bm{x}^\prime$ is

\begin{equation}
  I_\gamma\left(\bm{x},\bm{x}^\prime\right) = \frac{\kappa N_{\gamma}}{4\pi \left|\bm{x}-\bm{x}^\prime\right|^2}\exp\left(-\kappa \left|\bm{x}-\bm{x}^\prime\right|\right),
\end{equation}
where $\kappa^{-1}$ is the photon mean absorption length.
Assuming an infinite planar front, and evaluating the expression at a distance $z$ from the front yields

\begin{equation}
  \label{eq:Igamma}  
  \begin{split}
    I_\gamma(z) &= \sum_{i,j=-\infty}^{\infty}\frac{\kappa N_{\gamma}}{4\pi\left|z^2 + \Delta x^2\left(i^2 + j^2\right)\right|} \\
    &\times \exp\left(-\kappa \sqrt{z^2 + \Delta x^2\left(i^2 + j^2\right)}\right).
  \end{split}
\end{equation}

The number of physical photons generated per grid cell and time step in our simulations is typically $N_{\gamma} \sim \numrange{E2}{E9}$, depending on the location of the cell.
Since the plasma density in the streamer decays with distance from the streamer head, the center of the halo streamer head emits many more photons than the front.
The vertical length of the halo streamer head is also quite large, approximately \SI{1}{\kilo\meter}, so the photons emitted from the center are absorbed quite rapidly, on average about \SI{10}{\meter} from the emission site. 
Secondary electrons arising from these photons are therefore superimposed onto an already quite high plasma density, so these photons do not generate much preionization nor density fluctuations ahead of the halo.
Fewer ionizing photons are emitted from the leading edge of the halo streamer head due to the much lower plasma density in those regions.
An order-of-magnitude estimate from our simulations is $<\num{E3}$ ionizing photons emitted per grid cell and time step.
Unlike the photons that are emitted from the center of the streamer head, these photons escape the streamer head and generate pre-ionization in regions where the plasma density is initially low.
Thus, despite the fact that an overall large number of photons are emitted from the halo streamer head, a comparatively low number of photons are responsible for generating the electron density fluctuations at the halo front.

\begin{figure}[htb]
  \centering
  \includegraphics{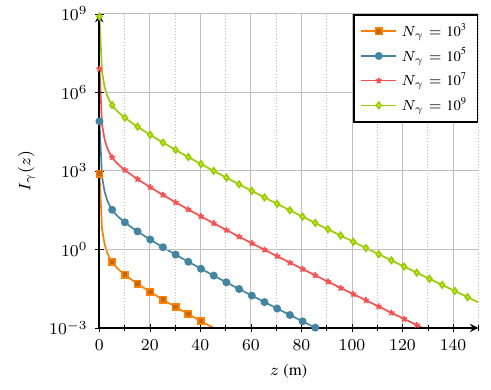}
  \caption{Radiative density as a function of distance $z$ from a planar front, calculated using equation~\eref{eq:Igamma}.
    Various numbers of initial photons $N_{\gamma}$ are included, using an absorption coefficient $\kappa = \SI[per-mode=reciprocal]{E-1}{\per\meter}$. 
  }
  \label{fig:DiscretePhotons}
\end{figure}

\Fref{fig:DiscretePhotons} shows the average number of photons absorbed per unit volume at various distances from the planar front in \fref{fig:PlanarFront}, using equation~\eref{eq:Igamma}.
For the curve with $N_{\gamma} \sim \num{E9}$ we find that single-photon effects manifest after $z\approx \SI{110}{\meter}$, which is about a factor of ten shorter than the width of the ionization layer of the halo.
For the curve with $N_{\gamma}\sim \num{E3}$, which is representative of the emissions from the leading edge of the halo, single-photon effects manifest after just $z\lesssim \SI{5}{\meter}$.
Thus, the discrete nature of photons manifests physically in the ionization zone of the halo.
Correspondingly, and in analogy with past suggestions \cite{Kulikovsky2000,Liu2004}, we conclude that the halo breakup is due to lack of sufficient photoionization for smoothing out plasma irregularities in the ionization zone, which ultimately trigger branching instabilities at the halo front.

Continuum models for photoionization \cite{Bourdon2007,Luque2007}, which are quite popular in the streamer simulation community, leads to suppression of halo front fluctuations.
\textcite{Bagheri_2019} and \textcite{Marskar2020} have quite conclusively demonstrated that continuum models are unsuitable when describing branching instabilities for laboratory streamers, a conclusion which can now also be extended to sprites.

\subsection{Column glow}
\label{sec:ColumnGlow}
Next, we evaluate the appearance of the glowing trail in the initial sprite streamer shown in \fref{fig:HaloBreakup}.
Originally, column glows for sprites were termed afterglows, and believed to occur due to delayed chemical reactions \cite{Sentman2008}.
More recent computer simulations show that the emission is due to sustained excitation into the light-emitting 1PN2 state \cite{Luque2010,Liu2010}.

\begin{figure}[h!t!b!]
  \centering
  \includegraphics{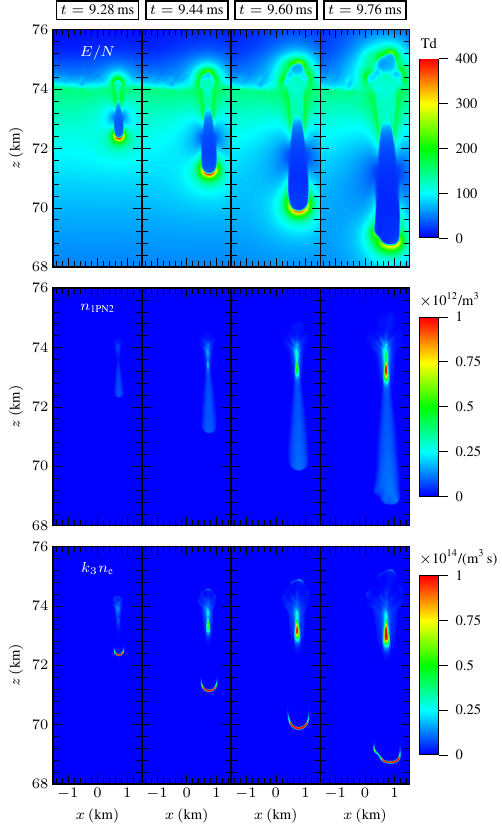}
  \caption{
    Snapshots of various simulation variables during glow formation (times are indicated above each column).
    Top row: Reduced electric field magnitude $E/N$.
    Middle row: Optical emission density $n_{\fpos}$.
    Bottom row: Dissociation term $k_3n_{\text{e}}$.
  }
  \label{fig:ColumnGlow}
\end{figure}

\Fref{fig:ColumnGlow} shows the quantitative development of the initial column glow for the streamer, where the data is taken through a cross-section of the filament shown in \fref{fig:HaloBreakup}.
In the three rows we show the reduced electric field magnitude $E/N$, the accumulated emissions $n_{\fpos}$, and the dissocative attachment term $k_3n_{\ch{e}}$.
We find that the glow develops just below the halo edge, and from a region where the electric field is initially quite high, approximately \SIrange{90}{120}{\townsend}.
From the row containing $n_{\fpos}$ we observe that the glow persists for a long time, and also extends downwards by about \SI{100}{\meter} from $t=\SI{9.28}{\milli\second}$ to $t=\SI{9.76}{\milli\second}$.
The row containing $k_3n_{\ch{e}}$ shows that electron attachment in the glow region simultaneously increases during the same time window.

\begin{figure}[h!t!b!]
  \centering
  \includegraphics{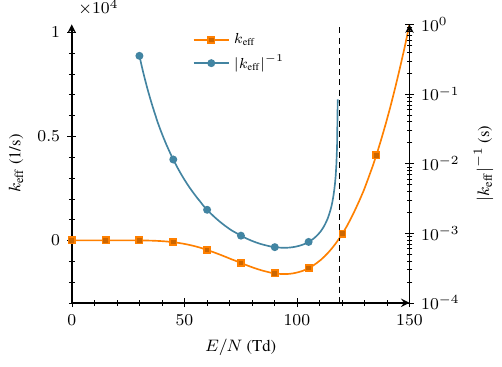}
  \caption{
    Effective ionization rate $k_{\text{eff}} = k_1 + k_2 - k_3$ at $z=\SI{73}{\kilo\meter}$, plotted against the left axis.
    The other curve shows $\left|k_{\text{eff}}\right|^{-1}$ for the attachment region $E/N \lesssim \SI{120}{\townsend}$ and is plotted against the right axis.
  }
  \label{fig:RateCoefficients}
\end{figure}

\textcite{Luque2016} propose that the column glow is due to an attachment instability \cite{DouglasHamilton1973} in the streamer channel.
While a detailed explanation can be found in \cite{Luque2016}, we consider a qualitative explanation of the phenomenon, which arises due to the nonlinearity of the impact ionization and attachment rates.
Because dissociative attachment turns electrons into slower-moving ions, the net effect of the dissociative attachment reaction $\ch{e} + \ch{O2} \rightarrow \ch{O^-} + \ch{O}$ is a lowering of the conductivity of the streamer channel. 
If, at some short segment in the streamer channel the rate of dissociative attachment is locally higher, the conductivity of that segment will also decrease faster than elsewhere.
Once the channel reaches a stationary or at the very least a quasi-stationary state, the electric field in that particular segment must correspondingly have increased due to charge conservation $\nabla\cdot\bm{J} = \nabla\cdot\left(\sigma\bm{E}\right)\approx 0$, where $\sigma$ is the conductivity and $\bm{J}$ is the current density.
Further evolution is then determined by the effective attachment rate, which is highly nonlinear in the electric field and is maximally effective around $E/N\approx\SI{90}{\townsend}$.
\Fref{fig:RateCoefficients} shows how the attachment rate $k_{\text{eff}} = k_1+k_2-k_3$ and the corresponding lifetime $\left|k_{\text{eff}}\right|^{-1}$ varies with $E/N$ at \SI{73}{\kilo\meter} altitude.
Once attachment begins to reduce the conductivity and thus increase the electric field, the process reinforces itself as the attachment rate grows further, and the process is mostly effective around $E/N\approx\SI{90}{\townsend}$.
Incidentally, this value coincides with the initial electric field in the upper part of the sprite in \fref{fig:ColumnGlow} where the effective attachment rate is $\left|k_{\text{eff}}\right|^{-1} \approx \SI{0.3}{\milli\second}$, so dissociative attachment in this region is fully capable of substantially reducing the conductivity of the channel during the formation time of the sprite.

The computer simulations show that the glow extends slightly downwards with time, and also extends into a region where the initial electric field was not high enough to effectively feed the attachment process.
For example, the rightmost column in \fref{fig:ColumnGlow} shows that the glow extends down to $z\approx\SI{72.5}{\kilo\meter}$, but the initial field at this altitude (top left panel in \fref{fig:ColumnGlow}) was $E/N\approx\SI{30}{\townsend}$, in which case the attachment lifetime is $\mathcal{O}\left(\SI{100}{\milli\second}\right)$.
This indicates that the column glow is not necessarily stationary, nor is there a hard cut-off into stable and unstable glow states as originally suggested by \textcite{Luque2016}.
On the other hand, our result suggest that the attachment process, conductivity reduction, and transition to a glow state is a continuous process that occurs everywhere in the channel at once.
The process just happens to be particularly effective around $E/N\sim \SI{90}{\townsend}$, and the corresponding decrease in conductivity rapidly leads to an increased electric field which then pushes the plasma towards a light-emitting state.
The optical emission from the glow then occurs as usual: Excitation and emission from the first positive system of nitrogen.

According to \fref{fig:RateCoefficients}, an initial electric field of $E/N$ somewhere between \SIrange{70}{110}{\townsend} is probably required in order to efficiently feed the attachment instability at the millisecond time scale (at \SI{73}{\kilo\meter} altitude).
Typical fields in streamer channels are usually much lower, so transition to glow states do not occur at observational time scales in conventional, unperturbed streamer channels.
For example, at around $E/N=\SI{20}{\townsend}$, which is an often observed field in positive streamer channels, the effective attachment lifetime at \SI{73}{\kilo\meter} altitude is about \SI{40}{\second}.
Finally, since the curves in \fref{fig:ColumnGlow} scale with $N$, analogous phenomena probably exist for atmospheric pressure streamers.
The mysterious pilot structures reported by \textcite{Kochkin2016b} could be indicative of an attachment-induced glow like we discuss above, but in this case it is not clear how the sufficiently high fields in the channel are initially reached.
One may also conjecture that such structures are prone to increased local gas heating since, at least to a coarse approximation, the local heat input is given as $\bm{E}\cdot\bm{J}$ but only $\bm{J}$ is constant in stationary channels.

\begin{figure*}[htb]
  \centering
  \includegraphics{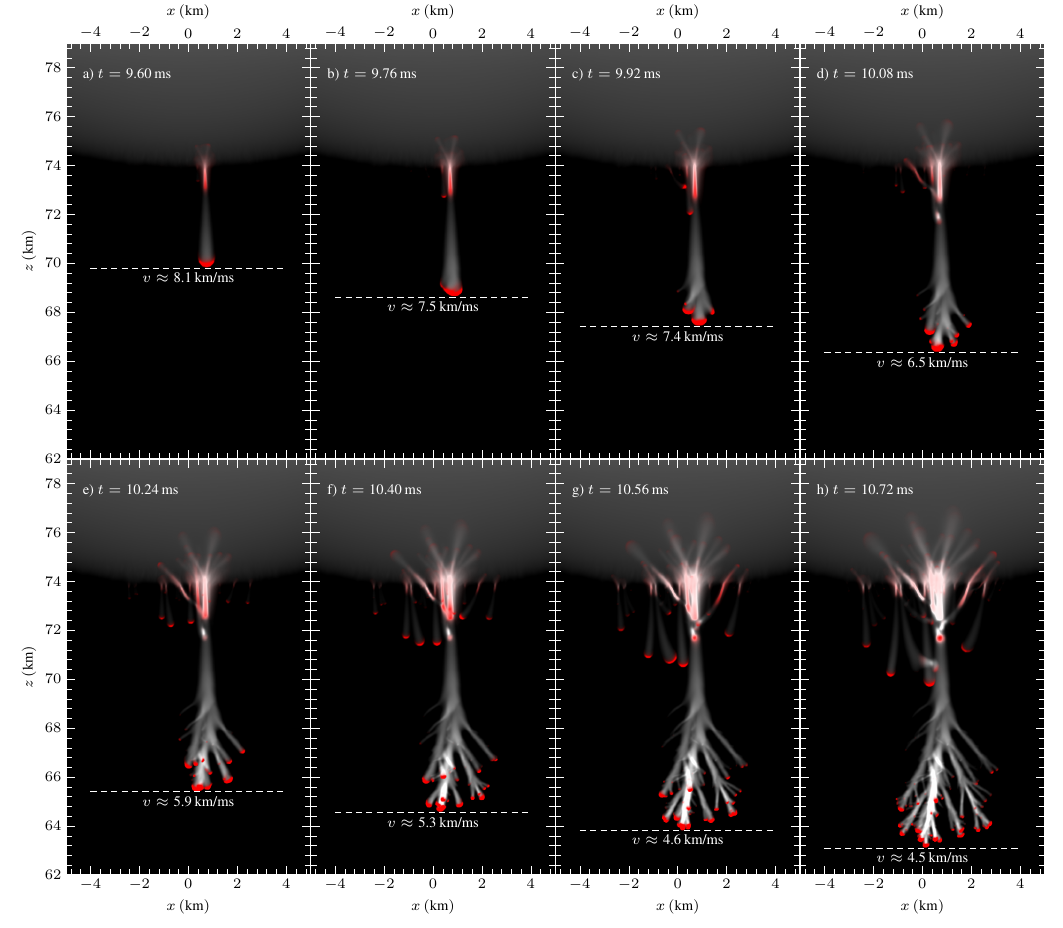}
  \caption{
    Long and short exposure images during the propagation phase.
    The dashed lines show the vertical streamer head position which were used to extract the indicated instantaneous velocities.
  }
  \label{fig:DownwardsPropagation}
\end{figure*}

Questions have been raised regarding the role of electron detachment from \ch{O^-} in the column glow.
\textcite{Kossyi1992} claim that detachment from \ch{O^-} by collisions with ground states of \ch{O2} and \ch{N2} are negligible, and provide rates for collisions with ground state \ch{O2} as well as excited molecules, specifically $\ch{N2}\left(\ch{A^3}\Sigma_u^+\right)$, $\ch{N2}\left(\ch{B^3}\Pi_g\right)$, $\ch{O2}\left(\ch{a^1}\Pi_g\right)$, and $\ch{O_2}\left(\ch{b^1}\Pi_g^+\right)$.
The same rates are reproduced in \textcite{Sentman2008}.
If we consider, for the sake of simplicity since we do not solve for any of these excited states, that the density of $\ch{N2}\left(\ch{A^3}\Sigma_u^+\right)$ is about the same as for \ch{N2^+}, the lifetime of the detachment reaction $\ch{O^-} + \ch{N2}\left(\ch{A^3}\Sigma_u^+\right) \rightarrow \ch{e} + \ch{O} + \ch{N2}$ at the observed glow altitude is approximately $\mathcal{O}\left(\SI{E3}{\second}\right)$ when we use the rates given by \textcite{Kossyi1992}.
As this reaction is the fastests among the \ch{O^-} detachment reactions provided by \textcite{Kossyi1992}, detachment due to collisions with the other excited states are also negligible.
Likewise, \textcite{Kossyi1992} provide a rate of \SI{5E-21}{\cubic\meter\per\second} for the reaction $\ch{O^-} + \ch{O2}\rightarrow \ch{e} + \ch{O3}$, but the time scale for electron detachment at sprite altitudes is then $\mathcal{O}\left(\SI{1}{\second}\right)$, so none of these reactions appear to be relevant on the time scale of sprites.
On the other hand, \textcite{Luque2016} and \textcite{Malagon2020} report that column glows disappear in the computer simulations when the electron detachment reaction $\ch{O^-} + \ch{N2} \rightarrow \ch{e} + \ch{N_2O}$ is included in the reaction set (\ch{N2} indicates the ground state).
The rate used for this reaction traces back to experiments by \textcite{Rayment1978} who claim that ground states of \ch{N2} participate in the reaction, which is at odds with claims by \textcite{Kossyi1992}.
\textcite{Janalizadeh2020} recently analyzed the experimental setup used by \textcite{Rayment1978} and propose that vibrationally excited \ch{N2} polluted the experiments, and thus that the result is at least partially in error.
Pending new experiments that clarify the roles of ground states and vibrationally excited states of \ch{N2} in the detachment reaction $\ch{O^-} + \ch{N2}\rightarrow \ch{e} + \ch{N2O}$, we have omitted this reaction from our chemistry set.

\subsection{Downwards propagation and branching}
Next, we analyze the propagation of the initial filaments after breakup from the halo and as they descend downwards.
\Fref{fig:DownwardsPropagation} shows the accumulated and instantaneous optical emissions until the various time instants indicated in each panel.
The following features are observed:

\begin{enumerate}
\item The initial streamer that originates in the halo breakup propagates down to approximately \SIrange{68}{70}{\kilo\meter} altitude before it branches.
\item As the streamer propagates downwards, excitation into the 1PN2 state in the tail of the initial streamer persists, which is optically observed as a glowing column at altitudes \SIrange{72}{74}{\kilo\meter}.
  These glows were analyzed \sref{sec:ColumnGlow}.
\item The sprite streamer velocity decreases with altitude, and is reduced from \SI{8.1}{\kilo\meter\per\milli\second} at \SI{70}{\kilo\meter} altitude to \SI{4.5}{\kilo\meter\per\milli\second} at \SI{63}{\kilo\meter} altitude.
\item Multiple streamer reconnections between individual filaments occur.
  These are particularly visible in \fref{fig:DownwardsPropagation}h) but occur as early as \fref{fig:DownwardsPropagation}c).
\item A persistent, glowing bead region forms at $z\approx\SI{72}{\kilo\meter}$, due to one of the streamer reconnections.
\item Negative streamers start from the top of the channels left behind by the downward streamers and propagate upwards into the halo.
\end{enumerate}

Beads, reconnections, and upward streamers are analyzed in subsequent sections and we consider first the computational observations behind streamer branching.
This is a fundamental property of streamers, and while no comprehensive theory for the phenomenon exists, the phenomenon is qualitatively understood.
Physically, branching can only occur in the presence of local modification of the space charge layer around the streamer, e.g., through flattening or protrusions of some regions of it, or in the presence of density fluctuations, all of which can lead to additional local field maxima in the ionization zone.
Such perturbations on the space charge layer can cause irregular growth of the plasma density ahead of the streamer, and thus act as nuclei for new branches.
Curvature or density perturbations in the space charge layer is thus a necessary, but not sufficient, requirement for branching as other effects, e.g., electron diffusion or photoionization, may stabilize the front \cite{Arrayas2002}.
In laboratory experiments, streamers in air normally branch into two and occasionally three filaments \cite{Heijmans2013}, while imaging of sprites show that streamers may split into anywhere from 2 to about 10 filaments \cite{StenBaekNielsen2013}.
In gases with less photoionization, branching is more frequent \cite{Nijdam2010}.

\begin{figure}[htb]
  \centering
  \includegraphics{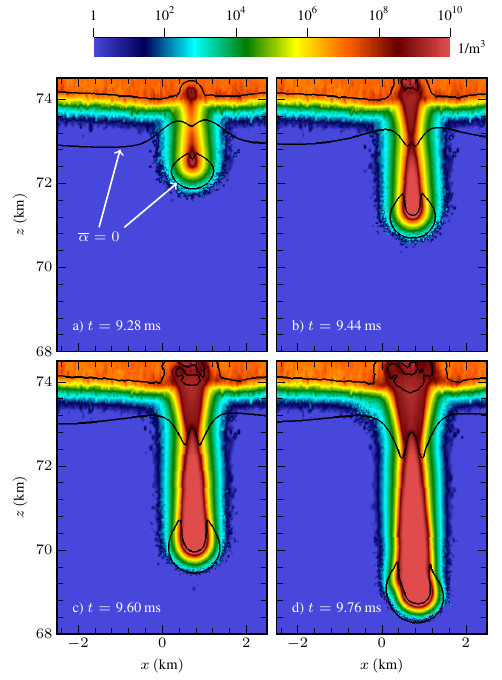}
  \caption{
    Time series of the electron density for the initial sprite streamer.
    The solid lines indicate the contour $\alpha=\eta$.
  }
  \label{fig:BranchingFluctuations}
\end{figure}

In the computer simulation we observe in \fref{fig:DownwardsPropagation}a) through c) that the sprite streamer radius increases as the streamer decends, and the ionizing photon absorption length simultaneously decreases.
From \fref{fig:AbsorptionLength} we find that at $z=\SI{80}{\kilo\meter}$ the mean absorption length of the photons is about \SIrange{4}{240}{\meter}, while at $z=\SI{70}{\kilo\meter}$ the corresponding lengths are \SIrange{1}{50}{\meter}.
When the first streamer breaks out of the halo at $z\approx \SI{74}{\kilo\meter}$ its optical radius is approximately \SI{250}{\meter}, while the maximum mean absorption length for photons is \SI{200}{\meter}.
As the streamer descends to \SI{70}{\kilo\meter} its radius increases to \SI{1}{\kilo\meter}, while the maximum mean absorption length simultaneously decreases to \SI{50}{\meter}.
\Fref{fig:BranchingFluctuations} shows a cross section of the electron density on the streamer axis, and demonstrates how these effects affect the streamer.
In \fref{fig:BranchingFluctuations}a), some photoelectrons appear outside the ionization volume ahead of the streamer, while \fref{fig:BranchingFluctuations}b) through d) show that the number of photoionization events outside the ionization volume gradually decreases with altitude.

A few qualitative evaluations of the underlying processes can now be made in the context of both laboratory and sprite streamers.
Firstly, sprites generate more ionizing photons per electron impact ionization event than corresponding laboratory streamers since collisional quenching is negligible at sprite altitudes but not at ground pressure.
For laboratory streamers, the amount of photoionization has been shown to substantially influence the amount of branching \cite{Bagheri_2019, Marskar2020,Wang2023}.
However, as discussed in \sref{sec:HaloBreakup} the distribution of the ionizing photons also matters, and the sprite streamers observed in the computer simulations have large radii compared to the photon absorption length.
As the volume of the ionization zone ahead of the streamer also increases with streamer radius, most of the emitted photons in a sprite streamer are, relatively speaking, absorbed close to the streamer head.
As with the sprite halo, the radiative intensity drops off exponentially with distance from the streamer head, e.g. with a characteristic length scale of approximately \SI{25}{\meter} at $z=\SI{70}{\kilo\meter}$, so far fewer photons reach the outer edges of the ionization zone.
Although more electron-ion pairs from photoionization are generated closer to the streamer head, the ones that are generated further out appear in regions where the electron density is already low, and they multiply exponentially over a longer distance than the ones starting closer to the streamer head.
For positive streamers these electron-ion pairs can thus exacerbate fluctuations in the front.
Incidentally, we believe this is also one of the reasons why small-diameter laboratory streamers branch less frequently than large-diameter ones \cite{Briels2008}, as the ionization zone for small-diameter streamers is then also smaller and discrete photon effects primarily manifest outside of it.
In summary, we propose that branching for sprite streamers occurs in precise analogy with laboratory streamers in air, i.e., due to growth of a Laplacian instability at the front of the streamer.

\subsection{Sprite beads}
Sprite beads are luminous spots that appear in sprite streamer channels, and whose time scales indicate that they are not slowly moving or stagnant streamer heads.
These structures are consistently captured in high-speed imaging experiments \cite{McHarg2008, Cummer2006}, and are both stationary and long-lasting.
\textcite{Cummer2006} report on bead formation due to streamer reconnection, i.e. collisisions between streamer channels, and suggest that a number of beads appear from this process.
Other types of beads appear to form spontaneously, or at the very least without a clear observation of a streamer reconnection, and the mechanism leading to formation of these types beads is currently not known.
\textcite{Luque2016} propose that bead emissions are due to the same process as glows, i.e. due to an attachment instability in the streamer channel.

\begin{figure}[h!t!b!]
  \centering
  \includegraphics{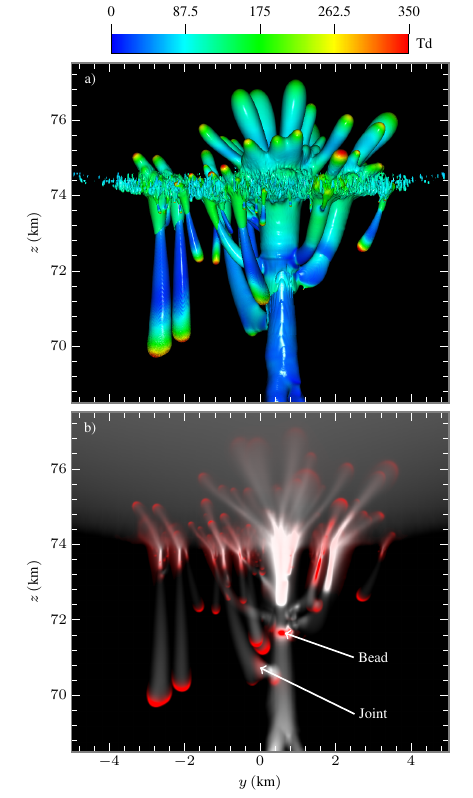}
  \caption{
    a) Isosurface plot of $n_{\fpos} = \SI[per-mode=reciprocal]{5E9}{\per\cubic\meter}$.
    Colors indicate values of the reduced electric field $E/N$.
    b) Image of accumulated and instantaneous sprite bead emissions.
    White colors indicate accumulated emission intensities and red colors indicate instantaneous emissions.
  }
  \label{fig:SpriteBead}
\end{figure}

In the computer simulation we observed numerous streamer reconnections, all of which occured between downward propagating streamers.
\Fref{fig:SpriteBead}a) shows an isosurface plot of the 1PN2 emissions (colors indicate electric fields), where several reconnections can be observed.
Note that the reconnections are due to late-emerging positive streamers propagating downwards from the halo edge and connecting to the tail of early streamers, and not upward shooting negative streamers.
Additional reconnections also occured, but are hidden behind some of the isosurfaces shown in the plot.
\Fref{fig:SpriteBead}b) shows a corresponding image of the accumulated and instantaneous optical emissions, where a bead has formed at $z\approx\SI{71.5}{\kilo\meter}$.

\begin{figure}[h!t!b!]
  \centering
  \includegraphics{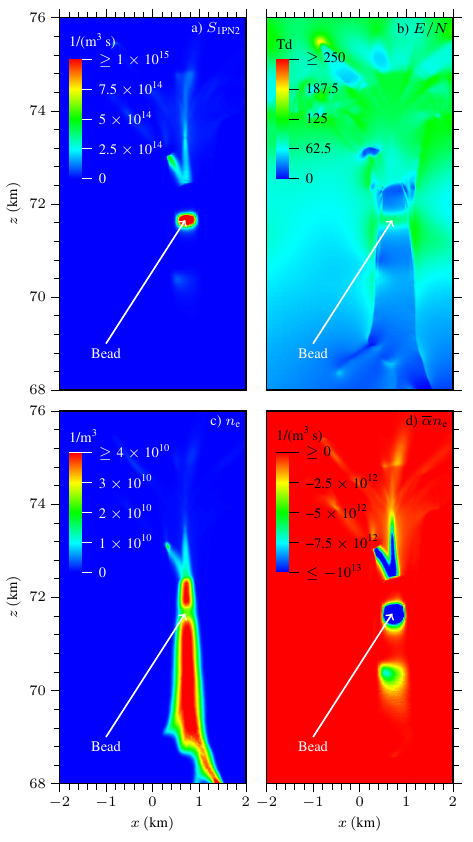}
  \caption{
    Two-dimensional slice through the streamer bead region at $t=\SI{10.72}{\milli\second}$.
    a) Instantaneous emissions from 1PN2.
    b) Reduced electric field.
    c) Electron density.
    d) Electron impact terms, i.e. $\overline{\alpha}n_{\text{e}}$.    
  }
  \label{fig:BeadSlice}
\end{figure}

To explain the emissive properties of the beads, \fref{fig:BeadSlice} shows a 2D slice of the simulation data through the streamer channel containing the bead indicated in \fref{fig:SpriteBead}.
In \fref{fig:BeadSlice} a) through c) we include the instantaneous emissions from 1PN2, the reduced electric field, and the electron density.
\Fref{fig:BeadSlice}d) also shows the attachment region of the electron impact term, i.e., the negative region of $\overline{\alpha}n_{\text{e}}$.
In the bead we observe the following features:
1) A localized spot of persistent 1PN2 excitations.
2) A localized and comparatively high electric field $E/N\approx\SI{90}{\townsend}$.
3) A locally decreased electron density, and thus also a locally decreased conductivity.
4) An increased rate of dissociative attachment $\ch{e} + \ch{O2} \rightarrow \ch{O^-} + \ch{O}$.
These features are the hallmark features of the attachment instability that we discussed in \sref{sec:ColumnGlow} and we thus arrive at the same conclusion as \textcite{Luque2016}: Beads are due to persistent excitation of the first positive system, fed by an attachment instability that locally increases the electric field and thus enhances optical emissions.

In the computer simulations, not all reconnections resulted in bead formation and the beads were often, but not always, centered on the streamer channels. 
In \fref{fig:SpriteBead} we have marked a section by 'Joint', and we have observed that these structures also cause persistent light emission, although the glow from these regions is much weaker than beads centered on the channel.
Additional manifestations of these glow features are available in \fref{fig:OtherBead} at the end of this paper.
Investigations of the electric field in these regions again display the typical signature of an attachment instability, and we thus conjecture that most (perhaps all) beads and long-lasting glowing structures are manifestations of the attachment instability.
However, bead formation only occurs at observational time scales if streamer channels are sufficiently perturbed by electric fields reaching approximately \SIrange{70}{110}{\townsend} in the channel, as this is the only range of electric fields where the attachment instability can be fed sufficiently fast.
Since the field is the primary trigger of the attachment instability, the evolution of sprite beads is quite subtle and there are probably several mechanisms that leads to their formation.
Our calculations show that beads occasionally appear in association with streamer reconnection, while spontaneous beads were not observed.
This is not to suggest that spontaneous bead emergence does not occur at all, but the conditions for it to happen were either not present in our computer simulations, or the simulation was not run long enough for these beads to appear.
In at least one case did we observe structures that could be interpreted as a spontaneous bead at lower altitudes ($z\approx\SI{69}{\kilo\meter}$ in \fref{fig:OtherBead}), but where closer examination of the data shows multiple streamer reconnections just above the bead region. 

\subsection{Sprite reconnection}

\begin{figure}[h!t!b!]
  \centering
  \includegraphics{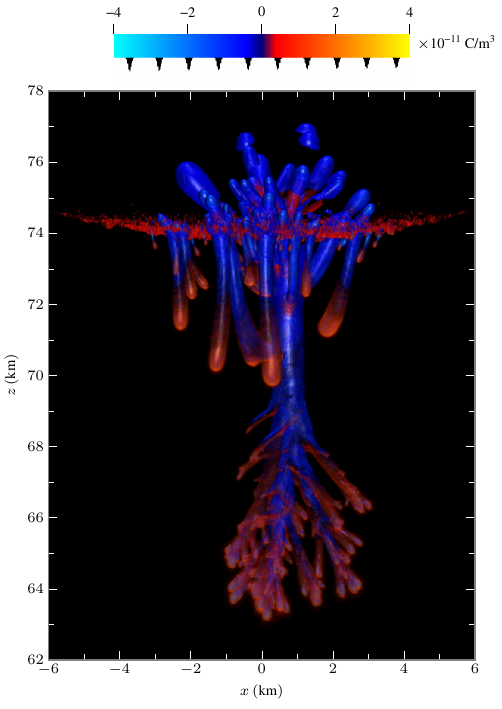}
  \caption{
    Space charge distribution $\rho$ in the sprite.
    Warm colors indicate positive space charge and cold colors indicate negative space charge, which can be mapped to the indicated color and opacities.
    The figure uses a truncated color range with semi-transparent layers so that the charge in both the core and space charge layers can be visualized.
    The curves at the bottom of the colorbar indicates which charge densities are visible.
  }
  \label{fig:SpaceChargeTree}
\end{figure}

For both sprite and laboratory streamers, various mechanisms for streamer reconnection \cite{Cummer2006, Nijdam2009} have been proposed.
\textcite{Luque2008} propose that under very specific circumstances merging can occur due to overlapping photoionization regions.
For this to happen, the streamers need to be separated by a distance shorter than the photon mean free path length, and the ionization volumes ahead of the streamers must also overlap.
However, in our computer simulation the streamers initiate quite far from each other, up to \SI{3}{\kilo\meter}, and still manage to reconnect, so merging of photoionization volumes cannot be the explanation in our simulations.
This type of reconnection has not been observed in other 3D computer simulations of laboratory streamers \cite{Marskar2020, Marskar2023Preprint} either.
As magnetic attraction is not represented in our computer simulations, and the late streamers bend into the main channel even though they are initially widely separated, the phenomenon must be purely electrostatic.
\textcite{Cummer2006} conjecture that this proceeds due to one streamer branching inducing a charge on the other, but as we will see, our calculations do not support this explanation either.
Rather, as first demonstrated by \textcite{Luque2010}, we find that negative charge accumulates on the tail of the streamers, which then attracts the positively charged heads of the late-emerging streamers.
The data corroborating this mechanism is given in \fref{fig:SpaceChargeTree}, which shows the space charge distribution of the sprites.
In the descending streamer heads the space charge distribution is reminiscent of laboratory streamers in the sense that they are slightly negatively charged in their core but have positively charged space charge layers.
However, both the core and outer layer of the tails of the sprite streamers are negatively charged, which is what causes the attraction between the early and late emerging streamers.

\subsection{Upwards propagation}
\Fref{fig:DownwardsPropagation} shows that several negative streamers also propagate upwards into the halo region.
All of the upward streamers that are visible in the figure start from the top of the channels left behind by downwards streamers, although some of these visually can appear as if they started from the bottom of the glow.
For example, the streamers on the right hand side of \fref{fig:SpriteBead}b) started as downwards streamers that connected to the main channel and then later transitioned into upwards streamers.

From \fref{fig:DownwardsPropagation}, \fref{fig:SpriteBead}, and \fref{fig:SpaceChargeTree} we extract the following qualitative information:
1) The upward streamers are characterized by negative charge in their heads and are thus negative streamers.
2) The upward streamers are characterized by a lower electric field at their tips than the downwards propagating positive streamers. Depending on their radius, we find fields roughly in the range \SIrange{150}{350}{\townsend}, while for the downwards positive streamers the fields are typically \SIrange{250}{450}{\townsend}.
3) There is a slight delay between the downward streamer initiation and the upward streamers. From \fref{fig:HaloBreakup} we observe that by the time the negative streamers initiate, the downward streamer has already propagated a few kilometers.
4) The luminousity of the upward streamers is comparatively low, and essentially merge with that of the halo.

As the upward streamers that appear in the computer simulations have low luminosity, we characterize our computer simulation as a column sprite rather than a carrot sprite, as the latter have very large luminous tops that can be easily identified \cite{McHarg2008}.
High-speed imaging of sprites show that upward streamers can be launched from the bottom of the column glow, which was recently demonstrated by \textcite{Malagon2020}.
This type of upwards propagation was not observed in our computer simulations.
However, returning to \fref{fig:BeadSlice}b) which shows the reduced electric field through a cross section of the streamer containing the bead, the electric field on the sides of the channel lies in the range \SIrange{90}{120}{\townsend}, which is just below the breakdown field.
These fields are found both around the bead region and the lower edge of the column glow, but clearly the fields were not sufficiently high for launching negative streamers.
Nonetheless, our simulations lends credence to the fundamental mechanism suggested by \textcite{Malagon2020}, as we here self-consistently show that comparatively high fields develop on the sides of the streamer channel in column glow and bead regions.

\section{Summary}
\label{sec:conclusion}

\subsection{Conclusions}

We have presented a theoretical examination of the birth and evolution of a column sprite in the Earth atmosphere, using self-consistent models and realistic atmospheric conditions.
The computer simulations allowed us to identify physical processes associated with commonly observed optical signatures.
We interpret the results as follows:
\begin{itemize}
\item
  Halo breakup occurs due to an instability at the halo edge, triggered by electron density fluctuations in front of the halo.
  These fluctuations originate due to absence of sufficient photoionization-enabled smoothing in the relatively long ionization zone of the halo as it descends into increasingly dense air.
\item 
  Sprite streamer branching occurs in analogy with both halo breakup; growing fluctuations at the streamer edge occur due to a lack of sufficient photoionization in the ionization zone of the streamers, which triggers branching.
\item
  The formation of a column-shaped glow region, which for the record is not a stationary structure, is attributed to the creation of a low-conductivity and high-field region within the streamer channels.
  This process is self-enhancing as the rate of dissociative attachment increases with the electric field magnitude up to $E/N\sim\SI{90}{\townsend}$, where the process is particularly fast and effective.
  In principle, this gradual transition towards a light-emitting state occurs everywhere in the streamer channels, but the process is primarily effective for $E/N\in\left[\SI{70}{\townsend},\SI{110}{\townsend}\right]$.
  The electric field in unperturbed channels is much lower than this, and such channels do not develop glows.
\item
  Optical emission from sprite beads, which have long perplexed researchers, also occur due to an attachment instability that is triggered by sprite streamer reconnection, in agreement with past suggestions \cite{Cummer2006, Luque2016}.
\item
  Streamer reconnection occurs due to electrostatic attraction between the positively charged heads of late emerging streamers and the negatively charged tails of early emerging streamers.
\end{itemize}

\subsection{Outlook}

\begin{figure}[h!t!b!]
  \centering
  \includegraphics{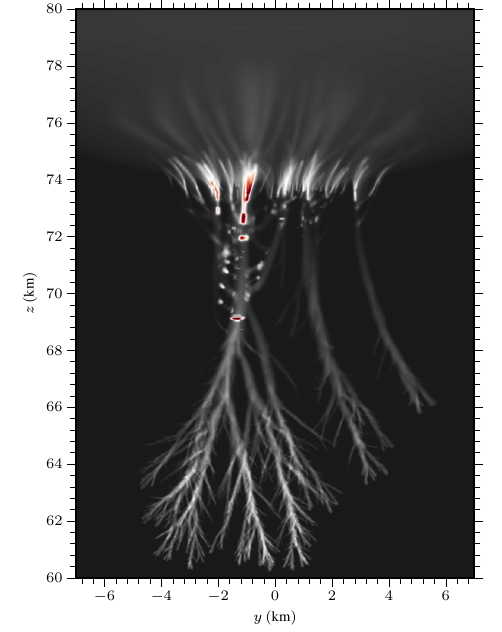}
  \caption{
    Example image of another computer simulation exposing a different sprite morphology in which multiple beads emerged, but sharing fundamental physical features with the in-depth analyzed sprite simulation.
    Note that the figure shows the accumulated sprite emissions but uses a different color map than preceding images.
  }
  \label{fig:OtherBead}
\end{figure}

In this paper we systematically analyzed a single sprite simulation, but remark that the above features emerged also in other computer simulations.
Like real sprites, there is great variability in the sprite morphology also in the computer simulations.
For example, although we selected a computer simulation in which a comparatively few number of branches emerged and only a single sprite bead was conclusively observed, other simulations displayed emergence of several beads. 
\Fref{fig:OtherBead} shows the accumulated optical emissions in a different computer simulation using precisely the same physical and numerical parameters, in which case additional beads can be identified.
These beads also developed due to the attachment instability, and the difference in number of beads is primarily related to the stochastic morphology of the sprite.

Our computer simulation(s) also explore a relatively narrow parameter space, as we do not explore different lightning strike characteristics nor atmospheric disturbances such as gravity waves \cite{Liu2015}, both of which are known to strongly influence the initiation and propagation of sprites.
Another issue is that precise ionospheric conditions are required in order to model specific sprites.
In this study we used a simplistic model for the ionosphere with a sharp decay below $z\approx \SI{83}{\kilo\meter}$, i.e., $L=\SI{1}{\kilo\meter}$ in equation~\eref{eq:init_ne}.
Recent calulations \cite{Kotovsky2016, Stenbaek-Nielsen2023} suggest that the electron density below this altitude could be higher than in our calculations.
However, as we discussed in \sref{sec:HaloBreakup} the free electrons are quite efficiently converted to negative ions ahead of the halo, and the level of preionization does not affect our conclusions regarding the basic formation mechanisms or optical structures. 
More specific ionospheric conditions can nonetheless affect the initiation altitude of the sprite, as well as the thundercloud charge required for initiation.

\subsection{Impact}

Our results hold dual significance:
First, we identify physical mechanisms associated with optical signatures, explaining previous high-speed imaging observations \cite{Cummer2006,McHarg2008}.
It is also clear that many sprite phenomena are linked and can not be understood in isolation.
For example, sprite reconnection occurs due to attraction between the negatively charged tail of a sprite and the positively charged head of another sprite, and the ensuing collision can then occasionally lead to apperance of sprite beads.
Second, we demonstrate the feasibility of performing comprehensive 3D simulations of sprites.
While this step required use of high-performance computing, these capabilities are crucial for quantifying the physical mechanisms of sprites, including the production of chemical species and their long-term impact on the Earth atmosphere.
Theoretical investigations of the type that is presented here are probably a key factor in achieving a complete quantitative understanding of sprites both on Earth and in other planetary atmospheres.

\section*{Acknowledgements}
This study was partially supported by funding from the Research Council of Norway through grant 321449.
The author expresses his gratitude to the Spritacular project \cite{Kosar2022} and Nicolas Escurat for supplying \fref{fig:RedSprite}.

\section*{Data availability statement}
The calculations presented in this paper were performed using the chombo-discharge computer code \cite{Marskar2023} (git hash 2456b1dd880).
Although the results included in this paper are stochastic, the input scripts containing simulation parameters and chemistry specifications that were used in this paper are available in both the chombo-discharge results repository at \url{https://github.com/chombo-discharge/discharge-papers/tree/main/Sprite}, and at the following URL/DOI: \url{https://doi.org/10.5281/zenodo.8434957}.

\printbibliography

\end{document}